\documentclass[11pt,preprintnumbers,aps,amssymb,nofootinbib,amsmath]
{revtex4}
\usepackage{epsfig,epsf}
\usepackage{graphicx}
\usepackage{verbatim}
\usepackage{epstopdf}
\usepackage{bm} % puts greek and math symbols in boldface using \bm
\newcommand{\beq}{\begin{equation}}
\newcommand{\beql}[1]{\begin{equation}\label{#1}}
\newcommand{\eeq}{\end{equation}}
\newcommand{\bea}{\begin{eqnarray}}
\newcommand{\eea}{\end{eqnarray}}

\def\eq#1{{(\ref{#1})}}
\def\fig#1{{Fig.~\ref{#1}}}

\def\sec#1{{Sec.~\ref{#1}}}

\newcommand{\as}{\alpha_s}
\newcommand{\bas}{\bar\alpha_s}

\newcommand{\Qg}{Q_\mathrm{geom}}

\def\b#1{\mathbf{#1}}

\begin{document}

\preprint{RBRC-740}

%~\vspace{1cm}

\title{ Probing the low-$x$ structure of  nuclear matter   with diffractive hadron production in pA collisions
%Diffractive hadron production in pA collisions at high energy
}

\author{Yang Li$\,^a$ and Kirill Tuchin$\,^{a,b}$\\}

\affiliation{
$^a\,$Department of Physics and Astronomy, Iowa State University, Ames, IA 50011\\
$^b\,$RIKEN BNL Research Center, Upton, NY 11973-5000\\}

\date{\today}


\begin{abstract}
We argue that  hadron production in coherent diffraction of proton on a heavy nucleus provides a very sensitive probe of the low-$x$ QCD dynamics. This process probes  the BFKL dynamics in proton and the non-linear gluon evolution in nucleus.  
 We calculate the diffractive hadron production  cross sections in the RHIC and LHC kinematic regions.  To study the nuclear effects we introduce the diffractive nuclear modification factor. We show 
that unlike the nuclear modification factor for inclusive hadron production that has a very interesting dynamics at RHIC but is expected to be almost completely saturated at the LHC, the nuclear modification factor  for diffractive production exhibits a non-trivial behavior both at RHIC and LHC.

\end{abstract}

\maketitle

%%%%%%%%%%%%%%%%%%%%%%%%%%%%%%%%%%%%%%%%
\section{Introduction}\label{sec:intr}

Over the last decade we have witnessed a remarkable success of models based on gluon saturation in description of low-$x$ data at HERA and RHIC. This allowed quantification of several key features of the low-$x$ dynamics of QCD. Still, there are many open questions such as the size of the NLO corrections to the BK equation, demarcation of boundary between the kinematic regions of gluon saturation and the collinear factorization, etc. These problems can be addressed by probing the nuclear structure at even smaller $x$  and/or by using a different set of measurements. In this paper we argue that diffractive hadron production in pA collisions is a measurement that can  provide a new handle on the low-$x$ nuclear dynamics.  Our study is  motivated  by the possibility to investigate the diffractive processes using the data on deuteron-gold (D-Au) collisions collected at RHIC.

A detailed theoretical analysis of coherent diffractive gluon production in onium-heavy nucleus ($q\bar qA$) collisions in the framework of the Color Glass Condensate \cite{Gribov:1983tu,Mueller:1986wy,McLerran:1993ni,Jalilian-Marian:1997dw,Kovner:2000pt,Iancu:2001md} was performed in our previous publications \cite{Li:2008bm,Li:2008jz}. There we argued that this process is  sensitive to the low-$x$ dynamics both in onium and nucleus. It has been argued in \cite{Avsar:2005iz,Avsar:2006jy,Avsar:2007xg} that it is phenomenologically  reasonable to approximate the proton light-cone wave-function (away from fragmentation regions) by a system of color dipoles. 
Additionally, we will demonstrate in \sec{sec:model} that the $qqqA$ propagator in the quasi-classical approximation takes exactly the same form as the $q\bar qA$ one if the emitted gluon transverse momentum is hard $k_T\gg Q_s$. In this approximation we can directly adopt the results of our theoretical analysis in Ref.~\cite{Li:2008bm,Li:2008jz}. The corresponding  phenomenological approach is developed in \sec{sec:nmf}. Similar model has been used in Ref.~\cite{Kopeliovich:2005us} for description of diffraction in pA collisions. 

There are several parameters that govern behavior of diffractive gluon production in pA collisions. These are:  gluon transverse momentum $k_T$ and rapidity $y$,   nucleus atomic number $A$ and transverse distance between the valence quarks $r_T$ in proton. The main observation of  \cite{Li:2008bm,Li:2008jz} is that  dependence of the diffractive hadron spectrum on these parameters in various  kinematic regions is quite different. This provides a convenient handle on the behavior of the low-$x$ gluon densities in the three most interesting kinematic regions: (i) gluon saturation region $k_T< Q_s$, (ii)  geometric scaling region $k_T<\Qg$ and (iii)  hard perturbative QCD region $k_T> \Qg$. 

The model that we use in this paper is based on analysis of diffractive hadron production in all available kinematic regions. Eq.~\eq{Isqr}  holds in the logarithmic approximation in all those regions and is therefore a convenient interpolation formula which we use to calculate the differential  inclusive cross section   \eq{main2}. The transverse vector $\b I(\b r', \b k, y)$ encodes information about the gluon density in the nucleus. It is related to  an integral of the forward elastic gluon dipole scattering amplitude $N_A(\b r', \b b, y)$ over all intermediate dipole size, see \eq{iq} and \eq{Q}. This amplitude is parameterized according to the KKT model \cite{Kharzeev:2004yx}. 
 On the other hand, the dipole density $n(\b r, \b r ', Y-y)$ encodes the gluon density in the proton, which is assumed to be dilute. Since the dipole density is a solution to the BFKL equation, we model it  by the LO BFKL amplitude in the diffusion approximation. 

To compare the low-$x$ dynamics in pA collisions to that in pp ones, it is convenient to introduce the \emph{diffractive} nuclear modification factor $R^{pA}_\mathrm{diff}$, see \eq{nmf}.  We evaluate the diffractive gluon production in pp collisions as a limit $A\to 1$ of that in pA ones. Theoretical expectations for $R^{pA}$ are detailed in \sec{nmfb} and \sec{nmfc}. The results of our numerical calculations performed using the KKT model \cite{Kharzeev:2004yx} are presented in \sec{nmfd}. We observe, that  $R^{pA}_\mathrm{diff}$ behavior is quite different from that of the nuclear modification factor $R^{pA}_\mathrm{incl}$ for inclusive hadron production. In the  RHIC kinematic region, at moderately large $k_T$ there is a significant enhancement of particle production in pA collisions, see \fig{G1}. This happens due to the fact that the diffractive cross section at large $k_T$ is proportional to the higher twist contribution that is enhanced in pA collisions by an additional factor of $ A^{1/3}$.  This enhancement gets increasingly compensated  at forward rapidities by  a suppression stemming from two sources: (i) gluon saturation in the nucleus; (ii) shrinking of phase space available for the  BFKL evolution in proton \cite{Li:2008bm}.
The latter feature of the diffractive hadron production is apparent in \eq{prot1} and is illustrated in \fig{F1} where we compare  $R^{pA}_\mathrm{diff}$ for two different diffusion coefficients (switching the BFKL evolution on and off).  In \fig{G2} and \fig{F2}  we show $R^{pA}_\mathrm{diff}$ at LHC. $R^{pA}_\mathrm{diff}$ exhibits 
rather strong dependence on rapidity. In contrast, $R^{pA}_\mathrm{incl}$ is not  expected to change a lot at LHC \cite{Tuchin:2007pf}. This implies, that by comparing  inclusive and diffractive hadron production in the wide kinematic region of RHIC and LHC one will be able to infer much of useful information about the higher twist contributions. 
Since different models of low-$x$ dynamics predict different dependence of higher twists  on atomic number $A$ and energy/rapidity, measurements of diffractive hadron production will be instrumental in determining the valid physical mechanism for hadron production at high energies.

%%%%%%%%%%%%%%%%%%%%%%%%%%%%%%%%%%%%%%%%
\section{A model for diffractive gluon production in pA collisions}\label{sec:model}

\subsection{Diffractive gluon production in $qqqA$ collisions}\label{sec:modelA}

\emph{Coherent} diffraction of a proton on a nucleus is a process $p+A\to X+A$ characterized by a large rapidity gap between the diffractive system $X$ and the \emph{intact} nucleus $A$. A fraction of the coherent diffractive events increases with the collision energy and is expected to reach its limiting value of a half at asymptotically high energies. In the mean-field approximation $\as\ll 1$ and $A\gg 1$, the 
  \emph{incoherent} diffractive processes such as  $p+A\to X+A^*$  where $A^*$ is a diffractive system of color-neutral nuclear debris, are parametrically suppressed. 
Therefore,  in the present paper we consider only the coherent diffraction\footnote{ The
  incoherent diffraction may be phenomenologically important at RHIC and LHC energies \cite{Kaidalov:2003vg,Kopeliovich:2005us}.}. 
  
Coherent diffraction is possible only if the coherence length $l_c$ of the emitted gluon with momentum $k$ is larger than the nucleus size $R_A$ (in the nucleus rest frame):
\beq\label{cL1}
l_c=\frac{k_+}{\b k^2}\gg R_A\,,
\eeq
where + indicates the light-cone direction of the incoming proton. The invariant mass of the produced system is given by $M^2=\b k^2/x$, where $x=k_+/p_+$ and $p$ is the proton momentum. Substituting these equations in \eq{cL1} yields  the following condition on the mass of the diffractive system:
\beq\label{cL2}
M^2\ll \frac{p_+}{R_A}=\frac{s}{R_Am_p}\,,
\eeq
where $\sqrt{s}$ is the center-of-mass energy of the proton--nucleon collision and $m_p$ is  proton mass.

\begin{figure}[ht]
      \includegraphics[width=8cm]{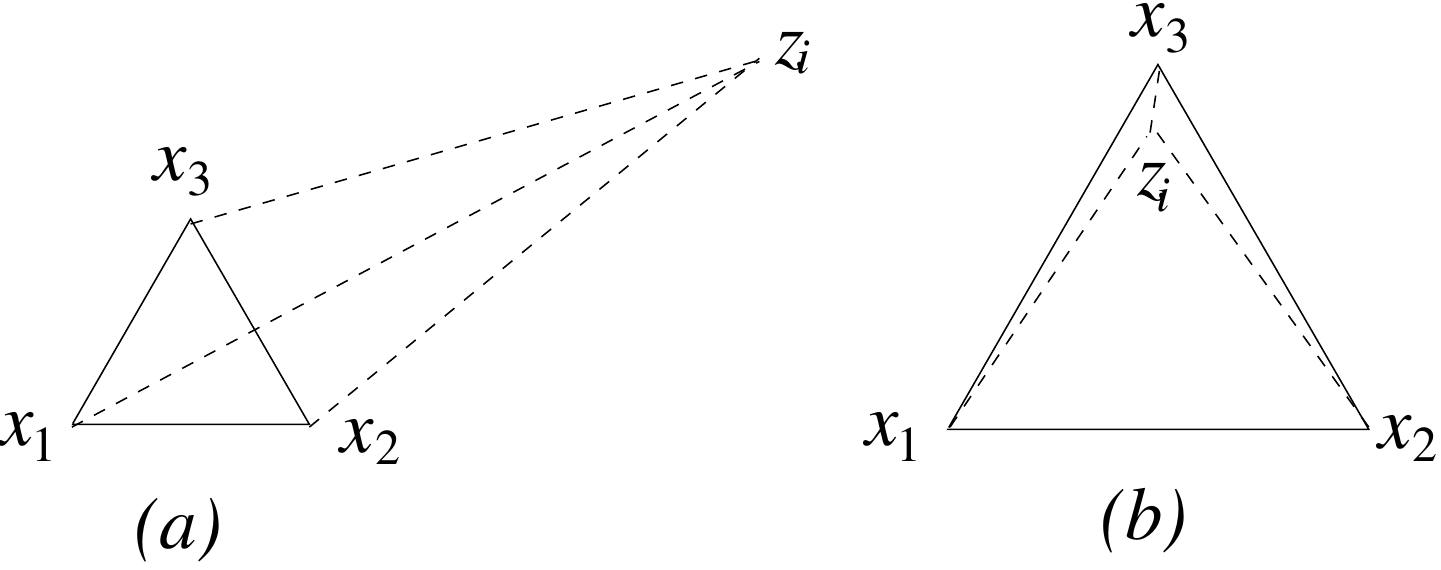}
\caption{Possible configurations formed by three valence quarks at positions $\b x_1$, $\b x_2$, $\b x_3$ and a gluon at position $\b z_i$. We assumed that $x_{12}\approx x_{23}\approx x_{31}$. In (a) sizes of the daughter dipoles are much bigger than sizes of the parent dipoles. For small enough $x_{12}$ this corresponds to the pQCD regime. In (b) one of the daughter dipoles is much smaller than the rest of dipoles corresponding to the high density regime (see  \cite{Levin:1999mw} for more details). }
\label{configs}
\end{figure}

A realistic model for diffractive gluon production in pA collisions was discussed by Kovchegov in \cite{Kovchegov:2001ni}. He considered, in the quasi-classical approximation, emission of a gluon by a color-neutral $qqq$ system of valence quarks with subsequent elastic interaction with a heavy nucleus. 
The resulting expressions for the propagators of the $qqq$ and $qqqG$ systems in the nucleus  can be written in the following form \cite{Kovchegov:2001ni}: 
\beql{piij}
\Pi_{ij}= v_i^T\left( e^{-M(\b z_1)}-e^{-(2/9)(\chi_{12}+\chi_{13}+\chi_{23})}\right)
\left( e^{-M(\b z_2)}-e^{-(2/9)(\chi_{12}+\chi_{13}+\chi_{23})}\right)v_j
\eeq
where 
\beql{vis}
v_1^T=(-1,\,\, -\frac{1}{\sqrt{3}}),\quad v_2^T=(1,\,\, -\frac{1}{\sqrt{3}}), \quad 
v_3^T=(0,\,\, \frac{2}{\sqrt{3}})\,,
\eeq
the $2\times2 $ matrix $M(\b z)$ is given by
\beql{matM}
M(\b z)= \left(\begin{array}{cc}
\frac{1}{6}\zeta_3+\frac{5}{12}(\zeta_2+\zeta_1)+\frac{5}{36}(\chi_{23}+\chi_{13})-\frac{1}{9}\chi_{12} & 
\frac{1}{4\sqrt{3}}(-\zeta_2+\zeta_1+\chi_{23}-\chi_{13})
\\
\frac{1}{4\sqrt{3}}(-\zeta_2+\zeta_1+\chi_{23}-\chi_{13}) & 
\frac{1}{2}\zeta_3+\frac{1}{4}(\zeta_2+\zeta_1)-\frac{1}{36}(\chi_{23}+\chi_{13})+\frac{2}{9}\chi_{12} 
\end{array}
\right)
\eeq
and the scattering amplitudes of various dipoles on a nucleon  read \footnote{In this section only we adopted a shorthand notation where the saturation scale is understood to include the logarithmic dependence on the dipole size. }
\beql{dist}
\zeta_i=\frac{1}{8}(\b z-\b x_i)^2Q_{s0}^2\,\quad \chi_{ij}=\frac{1}{8}(\b x_i-\b x_j)^2Q_{s0}^2\,,
\eeq
where $\b x_1$, $\b x_2$, $\b x_3$ are the valence quarks transverse coordinates, $\b z_1$ and $\b z_2$ are the gluon transverse coordinates in the amplitude and in the complex conjugated one respectively, see \fig{configs}.

The cross section for the diffractive gluon production in the quasi-classical approximation reads
\beql{xsectqqq}
\frac{d\sigma^{qqqA}}{d^2k_Tdy}= \frac{\as}{(2\pi)^2\pi^2}
\int d^2b\, d^2z_1 \, d^2z_2\,e^{-i\b k\cdot(\b z_1-\b z_2)}\,\sum_{i=1}^3\sum_{j=1}^3\, \frac{\b z_1-\b x_i}{|\b z_1-\b x_i|^2} \frac{\b z_2-\b x_j}{|\b z_2-\b x_j|^2}\, \Pi_{ij}\,.
\eeq

%%%%
%%%%%

Assume that the distances between the valence quarks  are approximately the same: $\chi_{ij}\approx \chi=\frac{3}{4}R_p^2Q_{s0}^2$, where $R_p$ is the proton radius. Then, matrix $M$ from \eq{matM} reduces to 
\beql{matMeq}
M(\b z)= \left(\begin{array}{cc}
\frac{1}{6}\zeta_3+\frac{5}{12}(\zeta_2+\zeta_1)+\frac{1}{6}\chi & 
\frac{1}{4\sqrt{3}}(-\zeta_2+\zeta_1)
\\
\frac{1}{4\sqrt{3}}(-\zeta_2+\zeta_1) & 
\frac{1}{2}\zeta_3+\frac{1}{4}(\zeta_2+\zeta_1)+\frac{1}{6}\chi
\end{array}
\right)\,.
\eeq
In general, the propagators $\Pi_{ij}$ are rather complicated objects. However,  in  the perturbative regime depicted in \fig{configs}(a)
 they can be reduced to a simple sum of the corresponding  $q\bar qG$ propagators as we are going to demonstrate now. The leading logarithmic  contribution in the perturbative regime stems from the configuration shown in \fig{configs}(a). In this case  $\zeta_i\gg \chi$ and we have $\zeta_1\approx \zeta_2\approx \zeta_3= \zeta$.  Eq.~\eq{matMeq} becomes
\beql{matMpert}
M(\b z)= \left(\begin{array}{cc}
\zeta & 0
\\
0 & \zeta
\end{array}
\right)\,.
\eeq
Using  \eq{piij}  and \eq{vis} we derive the propagator
\beql{propP}
\Pi_{ij}=
     \frac{4}{3}\left( e^{-\zeta(\b z_1)}-e^{-\frac{2}{3}\chi}\right)  
     \left( e^{-\zeta(\b z_2)}-e^{-\frac{2}{3}\chi}\right) \,(\delta_{ij}-\frac{1}{2}(1-\delta_{ij}))\,.                        
\eeq
The cross section \eq{xsectqqq} reads in this case 
\bea\label{a00}
&&\frac{d\sigma^{qqqA}}{d^2k_Tdy}\approx  \frac{\as\, }{(2\pi)^2\pi^2}
\int d^2b\, d^2z_1 \, d^2z_2\,
e^{-i\b k\cdot(\b z_1-\b z_2)}\nonumber\\
&&\qquad\times
 \frac{1}{2}\sum_{i< j}^3 \left( \frac{\b z_1-\b x_i}{|\b z_1-\b x_i|^2} -\frac{\b z_1-\b x_j}{|\b z_1-\b x_j|^2}\right)\cdot \left( \frac{\b z_2-\b x_i}{|\b z_2-\b x_i|^2} -\frac{\b z_2-\b x_j}{|\b z_2-\b x_j|^2}\right)\, \Pi_{11}\,.
\eea
In the 't Hooft's limit, Eq.~\eq{a00} can be related to the cross section for diffractive gluon production in quarkonium--nucleus collisions. To this end we introduce an effective color dipole with a quark and antiquark being at points $\tilde {\b x}_1$ and $\tilde {\b x}_2$ respectively. Then, in the same approximation as in \fig{configs}(a), we obtain 
\bea\label{xsectQ}
\frac{d\sigma^{q\bar qA}}{d^2k_T\, dy}&\approx&\frac{\as
C_F}{\pi^2}\frac{1}{(2\pi)^2}\,\int d^2b\, d^2z_1\,
d^2z_2\,\left(\frac{\b z_1-\tilde {\b x}_1}{|\b z_1-\tilde {\b x}_1|^2}-
 \frac{\b z_1-\tilde {\b x}_2}{|\b z_1-\tilde {\b x}_2|^2}\right)\cdot
 \left(\frac{\b z_2-\tilde {\b x}_1}{|\b z_2-\tilde {\b x}_1|^2}-
 \frac{\b z_2-\tilde {\b x}_2}{|\b z_2-\tilde {\b x}_2|^2}\right)\,\nonumber\\
 &&\times\,e^{-i\b k\cdot(\b z_1-\b z_2)}\,
 \left( e^{-\frac{1}{4}(\tilde {\b x}_1-\b z_1)^2\tilde Q_{s0}^2 }-e^{-\frac{1}{8}(\tilde {\b x}_1-\tilde {\b x}_2)^2\tilde Q_{s0}^2}\right)
  \left( e^{-\frac{1}{4}(\tilde {\b x}_1-\b z_2)^2\tilde Q_{s0}^2 }-e^{-\frac{1}{8}(\tilde {\b x}_1-\tilde {\b x}_2)^2\tilde Q_{s0}^2}\right)
  \,.
\eea
Hence
\beql{a02}
\frac{d\sigma^{qqqA}}{d^2k_Tdy}\approx \frac{2}{C_F}\frac{d\sigma^{q\bar qA}}{d^2k_Tdy}   =    \frac{3}{2}\frac{d\sigma^{q\bar qA}}{d^2k_Tdy}   \,.
\eeq
Comparing arguments of exponents in \eq{propP} and in \eq{xsectQ} we identify $\tilde Q_{s0}^2=\frac{1}{2}Q_{s0}^2$ as an effective saturation scale and 
\beql{a30}
\tilde R^2\equiv (\tilde {\b x}_1-\tilde {\b x}_2)^2=\, 2\cdot \frac{2}{3}\, ( {\b x}_1-{\b x}_2)^2=  \frac{4}{3}\cdot 3\, R_p^2 = (2\, R_p)^2
\eeq
as the square of the dipole  size. \footnote{In the following we are going to discuss only onium-nucleus scattering. Therefore we will omit the tildes 
to simplify notations.}
Expression \eq{a02} motivates a model that  we adopt in this paper.  We assume that the pA cross section can be approximated by $q\bar qA$ one with the dipole size given by \eq{a30}. This model correctly reproduces the pQCD limit. It also satisfies the unitarity bound, which is achieved in the saturation regime  depicted in \fig{configs}(b).

%%%%%%%%%%%%%%%
\subsection{Gluon production in quarkonium--heavy nucleus collisions}\label{sec2b}

%%%%
%\begin{figure}[ht]
 %     \includegraphics[height=8cm]{diffract1.pdf}
%\caption{Fan diagram describing the diffractive gluon production in pA collisions with  rapidity gap being equal to the rapidity of the produced gluon.}
%\label{fig:diffract1}
%\end{figure}
%%%%%

Now, as we set up a model for the diffractive gluon production in pA collisions in terms of the diffractive gluon production in $q\bar q A$ collisions, we would like to review the main results that we derived for the latter case in our previous publications \cite{Li:2008bm,Li:2008jz}. The cross section for the diffractive gluon production with transverse momentum $k_T$ at rapidity $y$ is given by 
\beq\label{main2}
\frac{d\sigma^{pA}(k_T,y)}{d^2k_Tdy} = \frac{\as C_F}{\pi^2}\frac{1}{(2\pi)^2}\,S_A\int d^2r' \, n_p(\b r, \b r', Y-y)\, |\b I(\b r',\b k,y)|^2\,,
\eeq
where $n_p(\b r, \b r', Y-y)$ is the dipole density in the projectile proton.
It  has the meaning of the number of dipoles of size $\b r'$ at rapidity $Y-y$  generated by evolution from the original dipole $\b r$ having rapidity $Y$ \cite{dip}. It satisfies the BFKL equation \cite{Kuraev:1977fs,Balitsky:1978ic} with the initial condition \eq{defnpic}.  The two-dimensional vector function $\b I(\b r',\b k,y)$ is defined as follows:
\beq\label{iq}
\b I(\b r',\b k,y)=-e^{-i\b k\cdot \b r'}\, i \nabla_{\b k}Q(\b r',\b k,y)+i\nabla_{\b k} Q^*(\b r',\b k,y)\,,
\eeq
where
\bea\label{Q}
&&Q(\b r', \b k, y)=-\int d^2w\, e^{i\b k\cdot \b w} \frac{1}{w^2}\nonumber\\
&&
\times \left[ N_A(\b r',\b b,y)-N_A(\b w-\b r',\b b,y)-N_A(\b w,\b b,y)+N_A(\b w-\b r',\b b,y)N_A(\b w,\b b,y)\right]\,.
\eea 
The vector function $\b I(\b r',\b k,y)$ incorporates information about two physical processes: (i) gluon emission off the daughter dipole $\b r'$ produced in the course of the BFKL evolution and (ii) low-$x$ gluon evolution in the nucleus through  
$N_A(\b r, \b b, y)$, which is the dipole-nucleus forward elastic scattering amplitude satisfying the BK equation \cite{Balitsky:1995ub,Kovchegov:1999yj}. In the quasi-classical approximation the dipole density reads
\beq\label{defnpic}
n_p(\b r,\b r',0)=\delta(\b r-\b r')\,,
\eeq
while the scattering amplitude is given by the Glauber-Mueller formula \cite{Mue} (now we explicitly write down the logarithm in the exponent) 
\beq\label{NQ}
N_A(\b r,\b b,0)= 1-e^{ -\frac{1}{8}\b r^2\,
 Q_{s0}^2\ln\frac{1}{r\Lambda}}\,,
\eeq
where $ Q_{s0}$ is the saturation scale at rapidity $y=0$.   In the case of dipole--proton scattering we expand \eq{NQ} and get 
\beql{npqc}
N_p(\b r, \b b, 0)=\frac{1}{8}\b r^2\,
\Lambda^2\ln\frac{1}{r\Lambda}\,.
\eeq

In all limiting cases we can write \cite{Li:2008jz}
\beql{Isqr}
|\b I(\b r',\b k,y)|^2\approx C\,\frac{4\,(2\pi)^2}{k^2}\, N_A^2(k^{-1}\hat {\b k}, \b b , y)\,[1-N_A(\b r', \b b , y)]^2\,
\sin^2\left( \frac{\b k\cdot \b r'}{2}\right)\,,
\eeq
where $C$ is a constant of order unity (its precise value  which can be found in \cite{Li:2008jz} is of little importance here). Let us emphasize, that \eq{Isqr} holds \emph{asymptotically} in \emph{all} kinematic regions. 
Due to the initial condition \eq{defnpic}, the cross section in the quasi-classical approximation is merely proportional to $|\b I(\b r',\b k,y)|^2$. Accordingly, employing \eq{Isqr} we obtain
\beql{xsectQC}
\frac{d\sigma^{pA}( R,k_T,0)}{d^2k_Tdy} \approx C\,\frac{4\,\as C_F}{\pi^2}\,\frac{S_A}{k_T^2}\, N_A^2(k_T^{-1}, \b b , 0)\,[1-N_A(R, \b b , 0)]^2\,
\sin^2\left( \frac{\b k\cdot \b R}{2}\right)\,,
\eeq

At larger rapidities we integrate over $\b r'$ in \eq{main2}
using \eq{Isqr}.  In the case of   hard gluons  we get 
\begin{eqnarray}\label{xssA}
&&\frac{d\sigma^{pA}( R,k_T,y)}{d^2k_Tdy}=\nonumber\\
&&\frac{\as C_F}{\pi^{5/2}}\, S_A\,N_A^2(k^{-1}, \b b , y)\,\frac{\min\{\frac{1}{k_T^2}, R^2\}\,}{\left(2\bas (Y-y)|\ln( Rk_T)|\right)^{1/4} }\,e^{2\sqrt{2\bas (Y-y)|\ln( Rk_T)|}}\,,
\quad k_T>Q_s\,,
\end{eqnarray}
where $\bas=\as N_c/\pi$. The cross section for the soft gluon production by a large dipole reads 
\beql{ivbsec2}
\frac{d\sigma^{pA}( R,k_T,y)}{d^2k_Tdy}=\frac{\as C_F}{8\pi^{5/2}}\,\frac{S_A}{Q_s^2}\,\frac{(2\bas (Y-y))^{1/4}}{\ln^{3/4}( RQ_s)}\,e^{2\sqrt{2\bas (Y-y)\ln( RQ_s)}}\,,\quad 
 R,\frac{1}{k_T}>\frac{1}{Q_s}\,,
\eeq
while in the case of soft gluon emission by a small onium 
\beql{ttt}
\frac{d\sigma^{pA}( R,k_T,y)}{d^2k_Tdy}= \frac{\as C_F}{4\pi^{5/2}}\,S_A\,  R^2\,\frac{1}{\left(2\bas (Y-y)\ln\frac{1}{ RQ_s}\right)^{1/4} }\,e^{2\sqrt{2\bas (Y-y)\ln\frac{1}{ RQ_s}}}\,,
\quad  R<\frac{1}{Q_s}<\frac{1}{k_T}\,.
\eeq
In all the reviewed cases \eq{xssA}-\eq{ttt} gluon multiplicity arises from the cut Pomeron that is hooked up to the incoming proton.

%%%%%%%%%%%%%%%
\subsection{Forward dipole--nucleus scattering amplitude}

The last required ingredient is the  forward elastic scattering amplitude $N_A(\b r,\b b, y)$. It can be evaluated in various kinematic regions.
In  the double logarithmic approximation  (DLA) 
\bea\label{naDLA}
N_A(\b r,\b b, y)=\frac{\sqrt{\pi}}{16\pi}\frac{\ln^{1/4}\left(\frac{1}{rQ_{s0}}\right)}{(2\bas y)^{3/4}}\,r^2Q_{s0}^2\,\left( 1+  \sqrt{\frac{2\bas y}{\ln\frac{1}{rQ_{s0}}}} \, \ln\frac{Q_{s0}}{\Lambda}  \right) e^{2\sqrt{2\bas y\ln\frac{1}{rQ_{s0}}}}\,,&&\nonumber\\
 r<1/Q_{s0}\,,\quad \ln\frac{1}{rQ_{s0}}\gg \as y\,.\qquad &&
\eea
This limit coincides with the small $x$ and small $r$ limit of the DGLAP equation. It obviously breaks the geometric scaling.  Consequently, the DLA  holds in the transition region between the gluon saturation and the hard perturbative QCD characterized by a hard scale $k_H$ , i.e.\ when $ Q_\mathrm{geom}<k_T<k_H$, where $Q_\mathrm{geom}$ is the scale at which the geometric scaling breaks down.  It reads  in the DLA
\beql{qgs}
Q_\mathrm{geom}\approx \frac{Q_s^2}{Q_{s0}}\,.
\eeq
The saturation scale is given by 
\beql{sat.s}
Q_s\approx A^{1/3}\Lambda^2\, e^{\lambda Y }\,,
\eeq
where $\lambda\approx 2\bas$ in the DLA.  
The hard scale $k_H$ can be related to the invariant mass of the diffractively produced system as discussed in Sec.~\ref{sec:modelA} in detail. 

As we approach the saturation region by decreasing $k_T$ at fixed rapidity we arrive at the diffusion approximation
\beq\label{naLLA}
N_A(\b r,\b b, y)=\frac{rQ_{s0}}{8\pi}\sqrt{\frac{\pi}{14\zeta(3)\bas y}}\ln\left(\frac{Q_{s0}}{\Lambda}\right) \, e^{(\alpha_P-1)y}\, e^{-\frac{\ln^2(rQ_{s0})}{14\zeta(3)\bas y}}\,,\quad \as y\gg \ln^2\left(\frac{1}{rQ_{s0}}\right)\,,
\eeq
where the BFKL Pomeron intercept is $\alpha_P-1=4\bas \ln 2$. We observe that the amplitude geometrically scales modulo small diffusive corrections.  The diffusion approximation \eq{naLLA} holds in the kinematic region $Q_s<k_T<Q_\mathrm{geom}$. 

Finally, when $k_T<Q_s$ solution to the BK equation deeply in the saturation region implies \cite{Levin:1999mw}
\beq\label{lt}
N_A(\b r, \b b, y)=1-S_0\, e^{-\tau^2/8}= 1-S_0\, e^{-\frac{1}{8}\ln^2(r^2Q_s^2)}\,,\quad r\gg \frac{1}{Q_s}\,.
\eeq

%%%%%%%%%%%%%%%%%%%%%%%%%%%%%%%%%
\section{Nuclear effects in diffractive gluon production} \label{sec:nmf}
%%%%%%%%%%%%%%

\subsection{Nuclear modification factor}

A convenient way to study the nuclear dependence of  particle production is to consider  the nuclear modification factor defined as follows
\beq\label{nmf}
R^{pA}_\mathrm{diff}(k_T,y)=\frac{\frac{d\sigma^{pA}_\mathrm{diff}(k_T,y)}{d^2k_Tdy}}{A\,\frac{d\sigma^{pp}_\mathrm{diff}(k_T,y)}{d^2k_Tdy} }\,.
\eeq
If the production process is completely incoherent then $R^{pA}_\mathrm{diff}(k_T,y)=1$. In the case of inclusive gluon production, the nuclear modification factor $R^{pA}_\mathrm{incl}$ was discussed in detail in \cite{Kharzeev:2003wz}. It has been demonstrated that in the extended geometric scaling region $Q_s(y)\lesssim k_T\lesssim Q_\mathrm{geom}$, the nuclear modification factor is suppressed as $R^{pA}_\mathrm{incl}\sim A^{-1/6}$, while in the saturation region $k_T\lesssim Q_s(y)$ the suppression is $R^{pA}_\mathrm{incl}\sim A^{-1/3}$. The amount of suppression is closely related to the value of the anomalous dimension $\gamma$ in a given kinematic region. At rapidity $y\simeq 0$ at RHIC $R^{pA}_\mathrm{incl}$ exhibits slight  enhancement (Cronin effect), which serves as indicator that the low-$x$ evolution in that process does not play an important role.  
We are going to argue below that the behavior of $R^{pA}_\mathrm{diff}$  is quite different from that of inclusive one which makes it a convenient tool for study of the low-$x$ gluon dynamics.

In the previous section we addressed in detail the diffractive gluon production in pA collisions.  In order to evaluate the $R^{pA}_\mathrm{diff}$ we need to normalize it by that in pp collisions. The latter is obtained by replacing the forward elastic dipole--nucleus scattering amplitude given by \eq{naDLA}, with the corresponding  forward elastic dipole--proton scattering amplitude 
\beql{protDLA}
N_p(\b r,\b b, y)=\frac{\sqrt{\pi}}{16\pi}\frac{\ln^{1/4}\left(\frac{1}{r\Lambda}\right)}{(2\bas y)^{3/4}}\,r^2\Lambda^2\, e^{2\sqrt{2\bas y\ln\frac{1}{r\Lambda}}}\,.
\eeq
In \eq{naDLA} we replaced $Q_{s0}$ by $\Lambda$ and set $A=1$.  
Since we assume that the gluon saturation effects are negligible in the proton, 
the cross section for the diffractive gluon production in pp collisions in the case of large characteristic proton size is obtained from \eq{xssA} by setting $A=1$ with the result
\beq\label{ppdif}
\frac{d\sigma^{pp}( R,k_T,y)}{d^2k_T\, dy}=\frac{\as C_F}{\pi^{5/2}}\,\min\left\{\frac{1}{k_T^2}, R^2\right\}\, S_p\,N_p^2(k_T^{-1}\hat {\b k}, \b b , y)\,\frac{1}{\left(2\bas (Y-y)|\ln( Rk_T)|\right)^{1/4} }\,e^{2\sqrt{2\bas (Y-y)|\ln( Rk_T)|}}\,.
\eeq
Similarly, we get in the quasi-classical approximation using \eq{xsectQC} 
\beql{prot1}
\frac{d\sigma^{pp}( R,k_T,0)}{d^2k_Tdy} \approx C\,\frac{4\,\as C_F}{\pi^2}\,\frac{S_p}{k_T^2}\, \frac{\Lambda^4}{64\,k_T^4}\,\ln^2\left(\frac{k_T}{\Lambda}\right)\,
\,e^{-\frac{1}{4} R^2\Lambda^2}\,
\frac{1}{2}(1-J_0( Rk_T))\,,
\eeq
where we averaged over the directions of the dipole $ {\b R}$ according to 
\beq\label{av}
\frac{1}{\pi}\int_0^\pi d\theta \, \sin^2\left( \frac{ 1}{2}k_T  R\cos\theta\right)=
\frac{1}{2}(1-J_0( Rk_T))\,.
\eeq
Gluon saturation effects in proton may be important at backward rapidities at the LHC. Taking them into account constitutes a difficult and not yet solved problem. Fortunately, effects associated with the gluon saturation in proton are not expected to significantly alter the nuclear dependence of our results since they are likely to cancel between the numerator and denominator of \eq{nmf}.

%%%%%%%%%%%%%%
\subsection{Quasi-classical approximation}\label{nmfb}

The nuclear modification factor in the quasi-classical approximation  and at high transverse momenta is derived by substitution of \eq{xsectQC} and \eq{prot1} into \eq{nmf} and deduce
\beq\label{j22}
R^{pA}_\mathrm{diff}(k_T,0)=A^{1/3}\left(1-\frac{1}{8} A^{1/3}\frac{\Lambda^2}{2\,k_T^2}\ln\frac{k_T}{\Lambda}\right)\,e^{-\frac{1}{4} R^2 Q_{s0}^2}\, , \quad k_T\gg Q_{s0}\,,
\eeq
where we take into account that $S_A=A^{2/3} S_p$ and $ Q_{s0}^2=A^{1/3}\, \Lambda^2$. According to \eq{j22} at very large $k_T$ and fixed $A$ the nuclear modification factor approaches a constant
\beq\label{j23}
R^{pA}_\mathrm{diff}(k_T, 0)\to A^{1/3}e^{-\frac{1}{4}A^{1/3}\ln A^{1/3}}\,, \quad k_T\to \infty\,.
\eeq
Eq.~\eq{j22} implies that  $R^{pA}_\mathrm{diff}(k_T,0)$ approaches unity from below as $k_T\to \infty$. In contrast to $R^{pA}_\mathrm{diff}(k_T,0)$, the nuclear modification factor for inclusive gluon production receives a positive 
power correction that  is a source of the Cronin enhancement observed in  inclusive gluon production in pA collisions.

\begin{figure}[ht]
  \includegraphics[width=10cm]{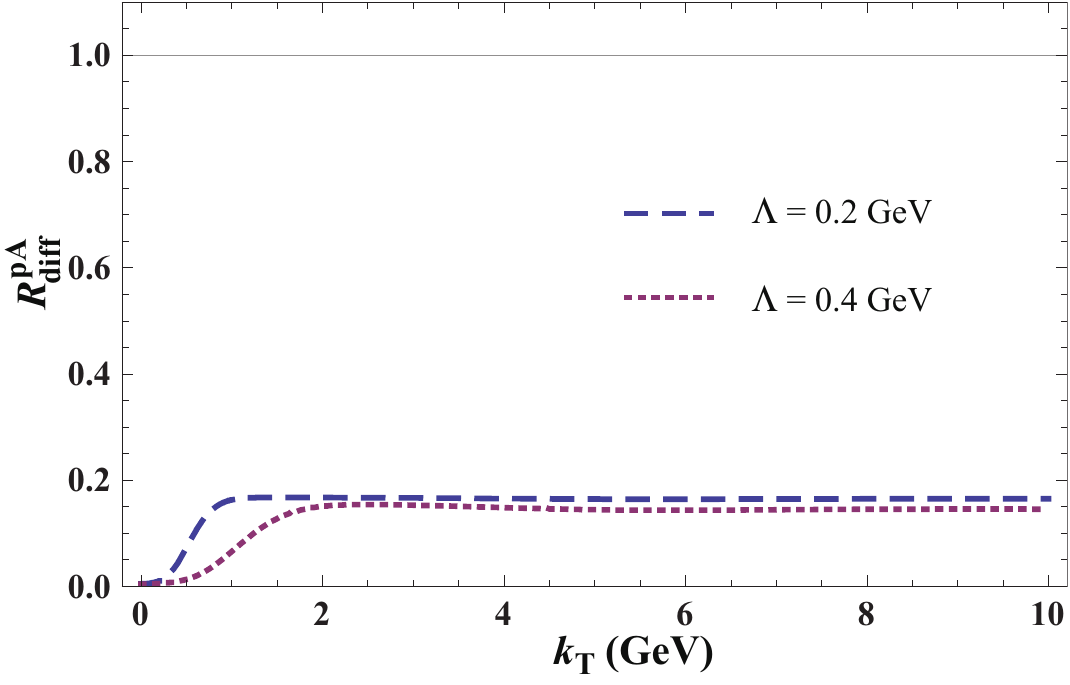}
\caption{Nuclear modification factor $R^{pA}_\mathrm{diff}$ as a function of transverse momentum $k_T$ in the quasi-classical approximation.  $\Lambda$ is a non-perturbative momentum scale.}
\label{fig:kT}
\end{figure}

In the saturation region we derive
\beq\label{j24}
R^{pA}_\mathrm{diff}(k_T,0)=\frac{64\, k_T^4}{A^{1/3}\, \Lambda^4\ln^2\frac{1}{ R\Lambda}}\,e^{-\frac{1}{4} R^2Q_{s0}^2}\,,
\quad k_T\ll  Q_{s0}\,.
\eeq
That is, the nuclear modification factor vanishes at small momenta as $k_T^4$. Actually, if we neglect the slow logarithmic dependence of the initial saturation scale $ Q_{s0}$ on $r$ in \eq{NQ} the integral appearing  in \eq{Q} can be taken analytically. The corresponding result can be found in \cite{Kovchegov:2001ni}. 
In \fig{fig:kT}  we use this analytical result to plot the nuclear modification factor $R^{pA}_\mathrm{diff}$ as a function of transverse momentum $k_T$. 

%\begin{figure}[ht]
 % \includegraphics[width=8cm]{Aspec.pdf}
%\caption{Nuclear modification factor $R^{pA}_\mathrm{diff}$ as a function of the atomic number $A$ in a quasi-classical approximation. $R^{pA}_\mathrm{diff}$ decreases monotonically with increasing $A$ and drops to around 0.1 for $A\sim 200$. The plot is drawn at $k_T=2$ GeV and $\Lambda=0.2$ GeV and is almost independent of the choice of $k_T$ and $\Lambda$.}
%\label{fig:A}
%\end{figure}

We observe that, unlike in the inclusive gluon production case, the size of the incoming projectile plays a very important role in the diffractive production. What is important is the relationship between the quarkonium size $ R$ and the inverse saturation scale $1/ Q_s$. In the quasi-classical approximation, that is neglecting the low-$x$ evolution, the diffractive gluon production is exponentially suppressed for heavy nuclei if $ R>1/ Q_s$ as compared to light nuclei. If $ R<1/ Q_s$ suppression gives way to enhancement at high transverse momenta. Both effects come about as  the result of the coherent scattering of proton off nucleus.

%%%%%%%%%%%%%%
\subsection{Low-$x$ evolution: hard gluons}\label{nmfc}
\subsubsection{Double logarithmic approximation}

The low-$x$ evolution has a dramatic effect on the diffractive gluon production. We would like to start our analyses with the case of  moderately large transverse momentum such that the geometric scaling is broken, but interaction is still coherent.  
Substituting \eq{xssA} and \eq{ppdif} into \eq{nmf} we derive that in general 
\beql{z1}
R^{pA}_\mathrm{diff}( R,k_T,y)=\frac{1}{A^{1/3}}\frac{N_A^2(k_T^{-1},\b b , y)}{N_p^2(k_T^{-1},\b b , y)}\,, \quad k_T\gg  Q_s\,.
\eeq
In the double logarithmic approximation, the BFKL equation coincides with the DGLAP equation. Therefore, in this region we can observe crossover from the coherent small-$x$ dynamics  to incoherent hard perturbative QCD. Using \eq{naDLA}  and \eq{protDLA} in \eq{z1} we derive
\beq\label{j1}
R^{pA}_\mathrm{diff}(k_T,y) = \frac{S_A}{A\, S_p}\,\sqrt{\frac{\ln\frac{k_T}{ Q_{s0}}}{\ln\frac{k_T}{\Lambda}}}\frac{ Q_{s0}^4}{\Lambda^4}\left( 1+\sqrt{\frac{2\bas y}{\ln\frac{k_T}{ Q_{s0}}}} \ln\frac{Q_{s0}}{\Lambda}\right)^2\,
e^{4\sqrt{2\bas y} \left(\sqrt{\ln \frac{k_T}{ Q_{s0}}   } - \sqrt{\ln\frac{k_T}{\Lambda}}  \right)   }\,,\quad k_T\gg Q_\mathrm{geom}\,.
\eeq
Introducing a new variable \cite{Kharzeev:2003wz}
\beq\label{j2}
\zeta= \left( \frac{\ln\frac{k_T}{ Q_{s0}}}{\ln \frac{k_T}{\Lambda}}\right)^{1/4} \,,
\eeq
we reduce \eq{j1} to
\beq\label{j3}
R^{pA}_\mathrm{diff}(k_T,y) =A^{1/3}\, \zeta^2\, \left( 1+\sqrt{2\bas y\,\ln\frac{Q_{s0}}{\Lambda}}\, \frac{\sqrt{1-\zeta^4}}{\zeta^2}\right)^2\exp\left\{
-4\sqrt{2\bas y \,\ln\frac{Q_{s0}}{\Lambda}}\sqrt{ \frac{1-\zeta^2}{1+\zeta^2}  }\right\}\,.
\eeq
The DLA approximation is valid when  $k_T\gg  Q_{s}(y)>  Q_{s0}$. In this case
 \eq{j3} becomes
\bea
R^{pA}_\mathrm{diff}(k_T,y)&\approx& A^{1/3} \left( 1-\sqrt{2\bas y} \frac{\ln\frac{Q_{s0}}{\Lambda}} {\sqrt{\ln\frac{k_T}{\Lambda}}} \right), \quad k_T \gg Q_{\rm geom} \, .\label{j5}
\eea
The remarkable feature of this result is enhancement of the nuclear modification factor by $A^{1/3}$. Unlike  the quasi-classical case \eq{j23}, this enhancement is not overrun at large $A$ by a small exponential factor.
The reason is that in course of low-$x$ BFKL evolution dipoles with small size $r<1/ Q_s$ are   produced and these dominate the cross section. Let us also mention that an enhancement similar to \eq{j5} has already been discussed in context of the $J/\psi$ production off the nuclear targets \cite{Kharzeev:2005zr} as well as in the breakdown of the collinear factorization of the fragmentation functions \cite{Li:2007zzc}.
 
It is important to emphasize that the result \eq{j5} holds only as long as the coherence length $l_c\approx \frac{1}{2M_Nx}$ is much larger than the nuclear size. Since in the center-of-mass frame kinematics $x=\frac{k_T}{\sqrt{s}}e^{-y}$, at large enough transverse momentum $k_T$ and fixed rapidity $y$ and energy $s$ the coherence is lost and the nuclear modification factor approaches unity. Therefore, the region where $R^{pA}_\mathrm{diff}\sim A^{1/3}$ scaling gives way to   
$R^{pA}_\mathrm{diff}\sim 1$ is the transition region between the semi-hard nuclear fields and the hard perturbative QCD. Needless to say that identification of this region is crucial for understanding the interplay between the dense and dilute high energy QCD regimes.

%%%%%%%%%%%%%%
\subsubsection{Extended geometric scaling region}

Next, we would like to analyze the extended geometric scaling region $Q_s(y)<k_T< Q_\mathrm{geom}$. Here the evolution is still linear and is well approximated by the leading twist approximation. However, the anomalous dimension of the gluon distribution significantly departs from unity and approaches the value it has at the critical line $k_T=Q_s(y)$. It is therefore appropriate to use the leading logarithmic approximation for the function $N_A(\b r, \b b, y)$. Substituting \eq{naLLA}  and \eq{protDLA} in \eq{z1}  we derive
\bea\label{j10}
R^{pA}_\mathrm{diff}(k_T,y) =\frac{4}{7\zeta(3)}\,\frac{k_T^2}{\Lambda^2}
\frac{\ln^2\left(\frac{Q_{s0}}{\Lambda}\right) \sqrt{2\bas y} }  {\sqrt{\ln\frac{k_T}{\Lambda}}}\,
\exp\left\{2(\alpha_P-1)y -4\sqrt{2\bas y\ln\frac{k_T}{\Lambda}} -\frac{2\ln^2\left(\frac{Q_{s0}}{k_T}\right)}{14\zeta(3)\bas y}    \right\}\,,&& \nonumber \\
Q_s<k_T< Q_\mathrm{geom}\,.\quad
\eea
This equation clearly demonstrates that the $A$-dependence of the nuclear modification factor arises only through the slow-varying logarithmic factors.
As far as the rapidity dependence is concerned, we can estimate it at the scale $k_T=Q_\mathrm{geom}(y)$.  Since  $N_A(\b r, \b b , y)$ is constant on the critical line  we derive
\beq
\label{j10-1}
R^{pA}_\mathrm{diff}(Q_\mathrm{geom}(y),y) \sim A^{1/3} \, e^{-4\sqrt{\bas \lambda }\,y}\,.
\eeq
That is, the nuclear modification factor is getting progressively suppressed in the forward direction.
This is much stronger suppression than in inclusive gluon production. Approximately  we can write
\beq\label{j12}
 R^{pA}_\mathrm{diff}(k_T,y)\sim A^{1/3}(R^{pA}_\mathrm{incl}(k_T,y))^2\,,\quad Q_s<k_T<Q_\mathrm{geom}\,.
\eeq
 Eq.~\eq{j12} clearly exhibits the higher twist nature of the diffractive gluon production. The peculiar properties of diffractive cross section due to the higher twist contributions in nuclear and hadronic DIS  have been discussed in \cite{Gotsman:2000zk,Gotsman:2000fy}.

%%%%%%%%%%%%%%
\subsubsection{Saturation region}
In the saturation region we utilize  \eq{ppdif} and one of the \eq{ivbsec2} or \eq{ttt} in \eq{nmf}  and arrive at a rather involved expression. Keeping only the parametric dependence and omitting the logarithmic factors we obtain 
\beq\label{j15}
R^{pA}_\mathrm{diff}(R,k_T,y)\sim \frac{1}{A^{1/3}}\frac{k_T^4}{ R^2\Lambda^4 Q_s^2}\,e^{2\sqrt{2\bas (Y-y)\ln( RQ_s)}}\,e^{-2\sqrt{2\bas (Y-y)\ln( Rk_T)}}\, e^{-4\sqrt{2\bas y \ln\frac{k_T}{\Lambda}}}\, 
\,  ,\quad  k_T,\frac{1}{R} < Q_s\,.
\eeq
There is a very strong suppression of diffractive gluon production in the saturation region in the case of low-$x$ evolution. This suppression however is still milder than in the quasi-classical case \eq{j24}. On the critical line $k_T=Q_s(y)$ we get for forward rapidities ($Y-y\ll y$) and central collisions (employing \eq{sat.s})
\beq\label{j20}
R^{pA}_\mathrm{diff}(Q_s(y),y)\sim \, e^{-4\sqrt{\bas \lambda }\,y}\,,
\eeq
which implies a strong suppression in the forward direction.

%%%%%%%%%%%%%%%
\section{Numerical calculations}\label{nmfd}

All the features that we discussed in the previous section can be visualized using a simple model for the forward elastic dipole--nucleus scattering amplitude $N_A(\b r, \b b, y)$. We parameterize it as follows
\cite{Kharzeev:2004yx}  
\beql{modN}
N_A(\b r,\b b ,y)=1-\exp\left\{ -\frac{1}{4}(r^2 Q_s^2)^{\gamma(r,y)}\right\}\,.
\eeq
 The anomalous dimension is parameterized in such a way that it satisfies the analytically well-known limits of (i) $r\to 0$, $y$ fixed and (ii) $y\to \infty$, $r$ fixed:
\beq\label{gamma} 
\gamma(r,y)=\bigg\{ \begin{array}{ccc}
     \frac{1}{2}\left(1+ \frac{\xi(r,y)}{|\xi(r,y)|+\sqrt{2|\xi(r,y)|}+28\zeta(3)} \right) & y\ge y_0\,,\\
     1        & y<y_0\,,
     \end{array}
\eeq
where 
\beq\label{xii}
\xi(r,y)=\frac{\ln\left[ 1/(r^2 Q_{s0}^2)\right]}{(\lambda/2)(y-y_0)}\,.
\eeq
In the double logarithmic approximation we can replace 
$ r^2\approx 1/(4  k_T^2)$.  The gluon saturation scale  is given by
\beq\label{satt}
Q_s^2(y)=\Lambda^2\, A^{1/3}\, e^{\lambda y} \left(\frac{\sqrt{s}}{200\,\mathrm{GeV}} \right)^\lambda\,,
\eeq
where parameters $\Lambda=0.6$ GeV and $\lambda=0.3$ are fixed by DIS data \cite{Golec-Biernat:1998js}. The initial saturation scale used in \eq{xii} is defined by $Q_{s0}^2=Q_s^2(y_0)$ with $y_0$ the value of rapidity at which the small-$x$ quantum evolution effects set in. Fit to the RHIC data yields $y_0=0.5$ \cite{Kharzeev:2004yx}. 

Numerical calculations of the cross section \eq{main2} are performed after substitution of  \eq{Isqr} with \eq{modN} and the following formula for the dipole density in diffusion approximation (cp.\ \eq{naLLA}): 
\beq\label{ndiff}
n_p( r, r', Y-y)=\frac{1}{2\pi^2}\frac{1}{rr'}\sqrt{\frac{\pi}{14\zeta(3)\bas\,d\, (Y-y)}}\, e^{(\alpha_P-1)(Y-y)}\, e^{-\frac{\ln^2\frac{r}{r'}}{14\zeta(3)\bas\, d\, (Y-y)}}\,.
\eeq
\begin{figure}[ht]
\begin{tabular}{cc}
  \includegraphics[width=8cm]{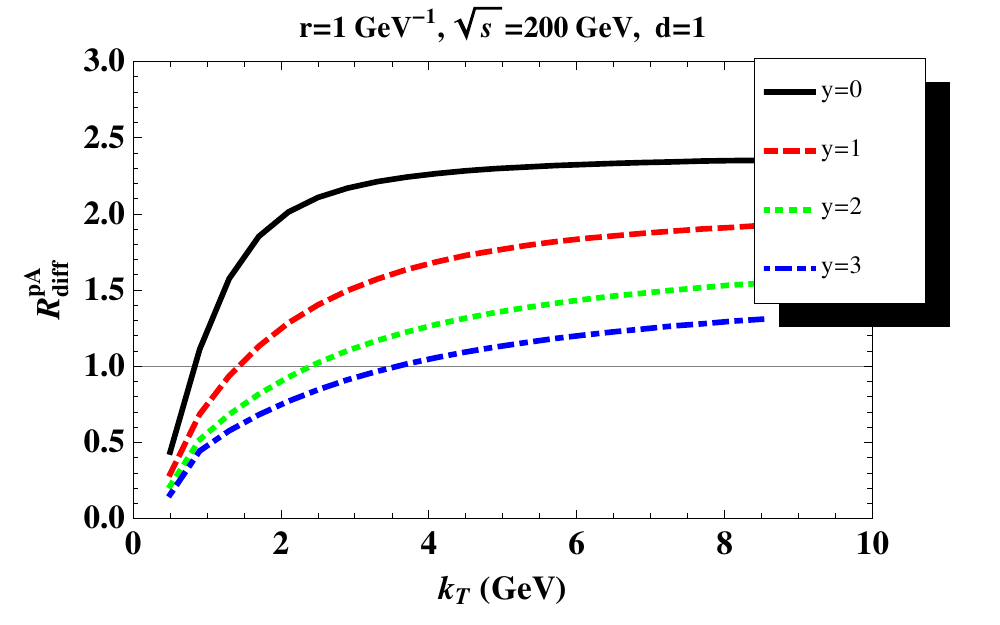}&
   \includegraphics[width=8cm]{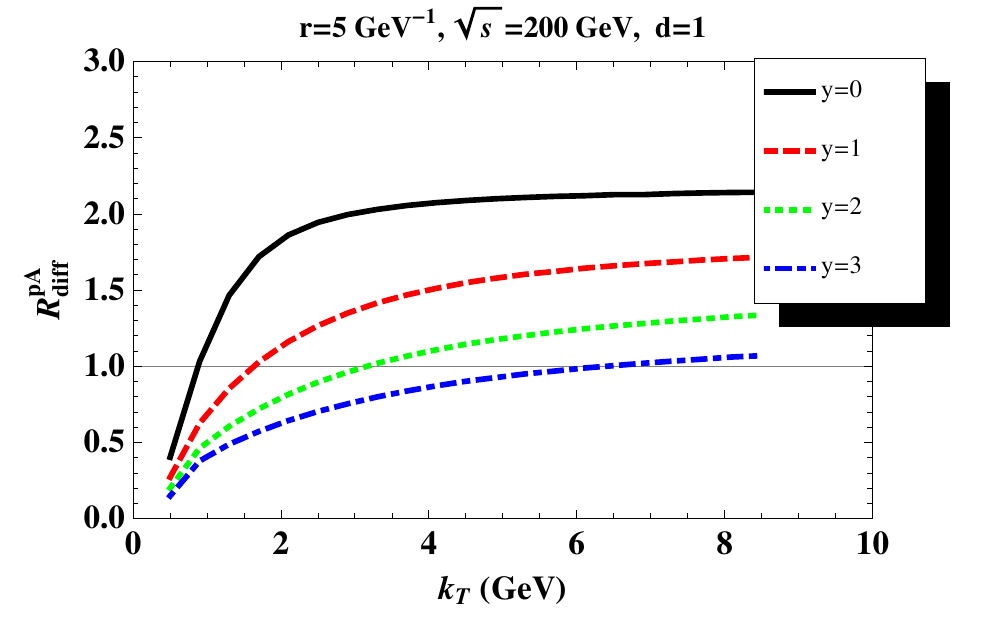}\\
   (a) & (b) 
   \end{tabular}
\caption{Nuclear modification factor for diffractive pion production in pA collisions at RHIC as a function of transverse momentum for two characteristic sizes of proton (a) 0.2 fm and (b) 1 fm. The effects of finite coherence length are neglected.}
\label{G1}
\end{figure}
Parameter $d$ is equal to unity in the LO BFKL. To obtain the hadron diffractive cross section we convoluted the obtained result with the LO pion fragmentation function given in \cite{Kniehl:2000hk}.   Diffractive gluon production in pp collisions, which is required as a baseline for the calculation of the nuclear modification factor \eq{nmf}, is obtained by setting $A=1$ in the formula for the corresponding cross section in pA collisions. 

\begin{figure}[ht]
\begin{tabular}{cc}
  \includegraphics[width=8cm]{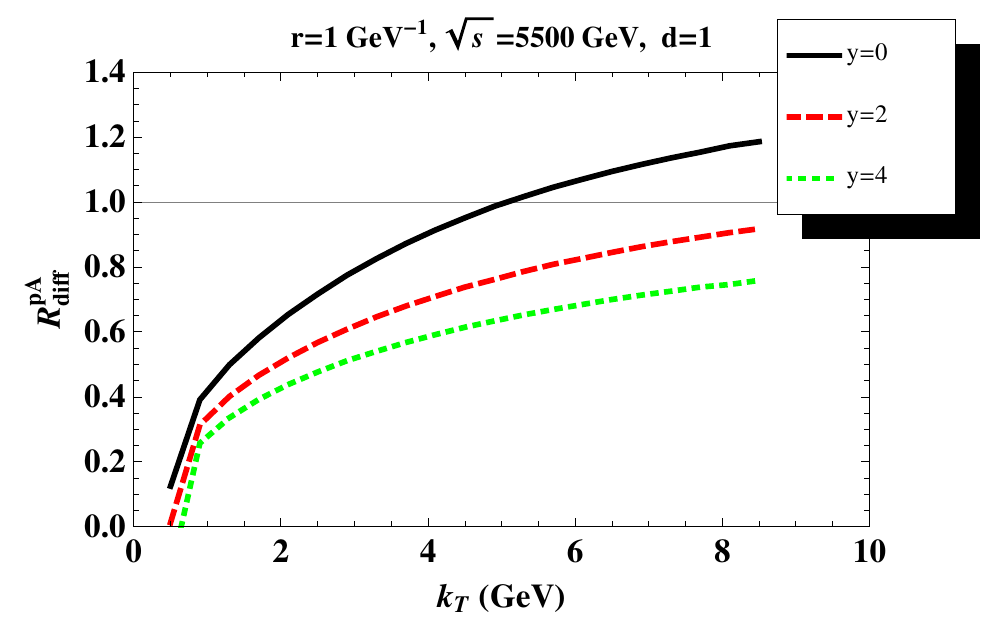}&
   \includegraphics[width=8cm]{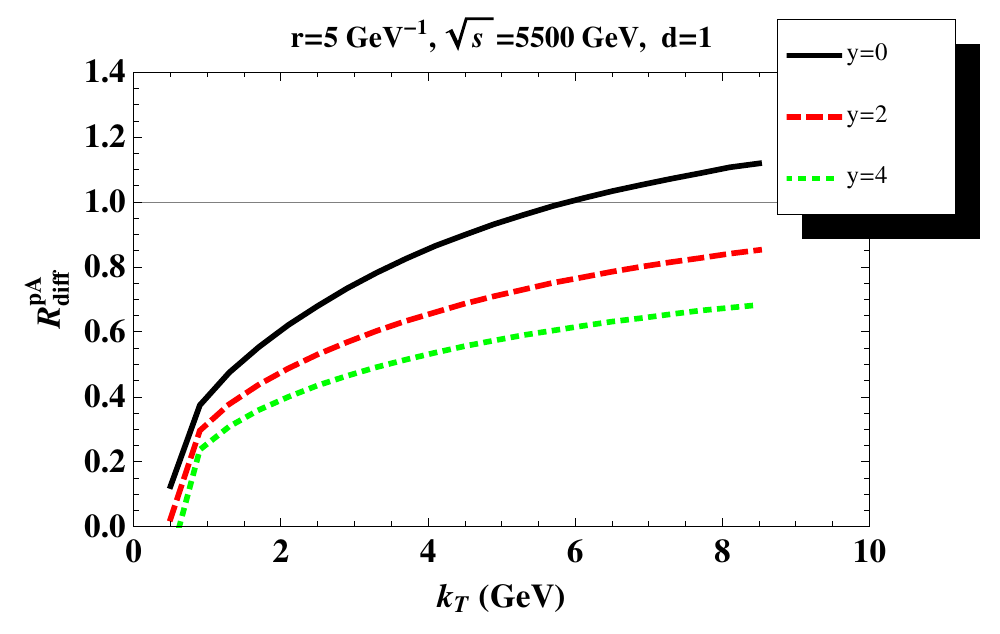}\\
   (a) & (b) 
   \end{tabular}
\caption{Nuclear modification factor for diffractive pion production in pA collisions at LHC as a function of transverse momentum for two characteristic sizes of proton (a) 0.2 fm and (b) 1 fm.  The effects of finite coherence length are neglected.}
\label{G2}
\end{figure}

\begin{figure}[ht]
\begin{tabular}{cc}
   \includegraphics[width=8cm]{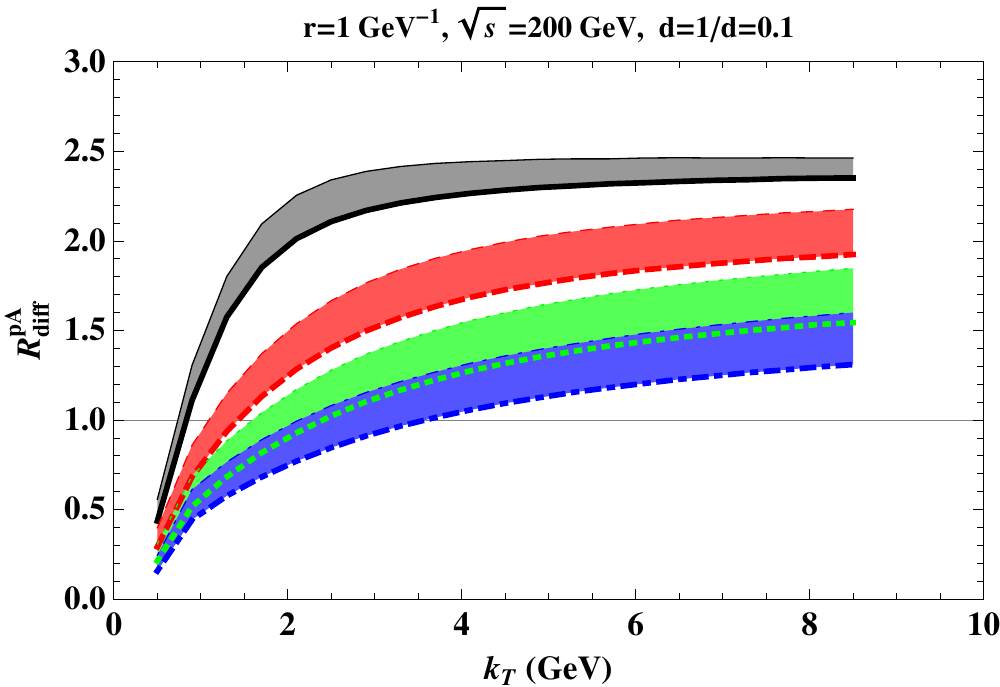}&
  \includegraphics[width=8cm]{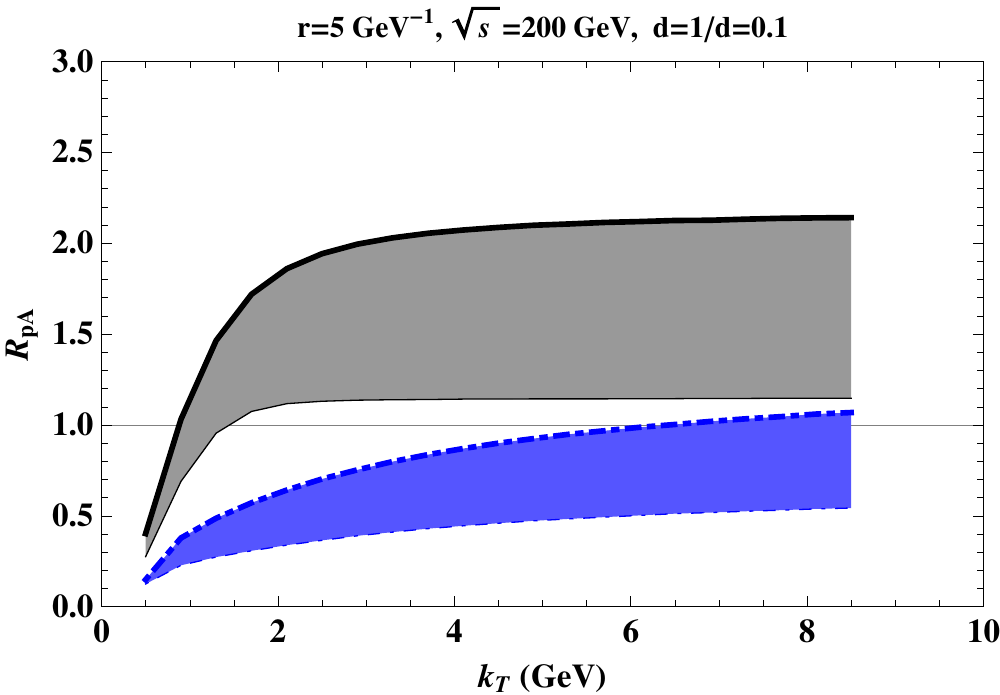}\\
   (a) & (b) 
\end{tabular}
\caption{Effect of diffusion in the dipole sizes on the diffractive pion production at RHIC for  two characteristic sizes of proton (a) 0.2 fm and (b) 1 fm. Upper line of the same type corresponds to $d=1$, the lower one -- $d=0.1$.  Lines of different types correspond to different rapidities (notations are the same as  in \fig{G1}). The effects of finite coherence length are neglected.}
\label{F1}
\end{figure}

\begin{figure}[ht]
\begin{tabular}{cc}
 \includegraphics[width=8cm]{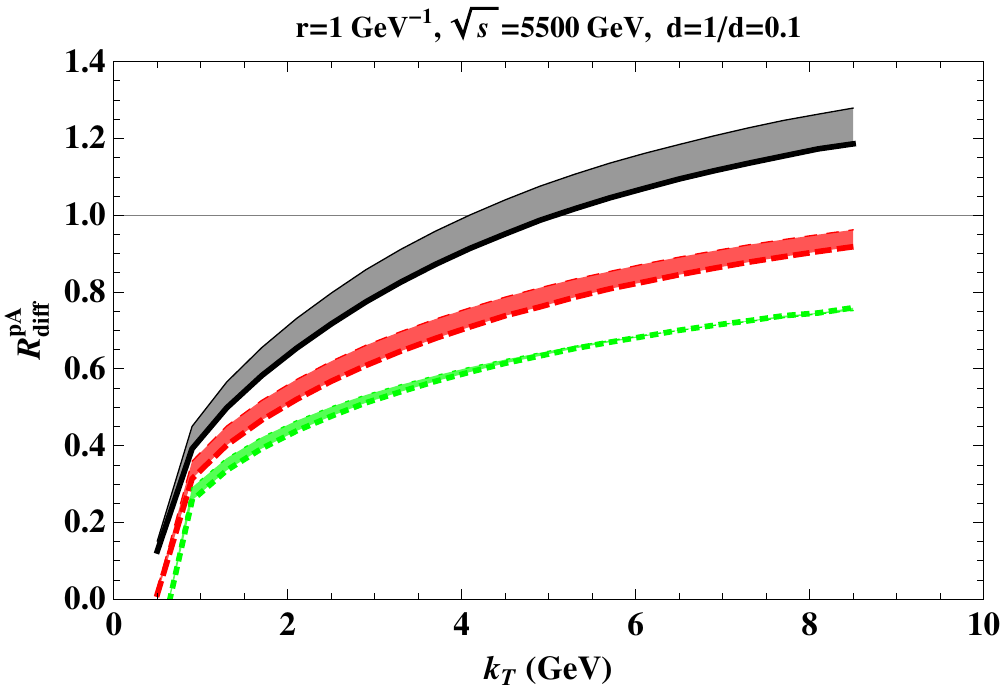}&
   \includegraphics[width=8cm]{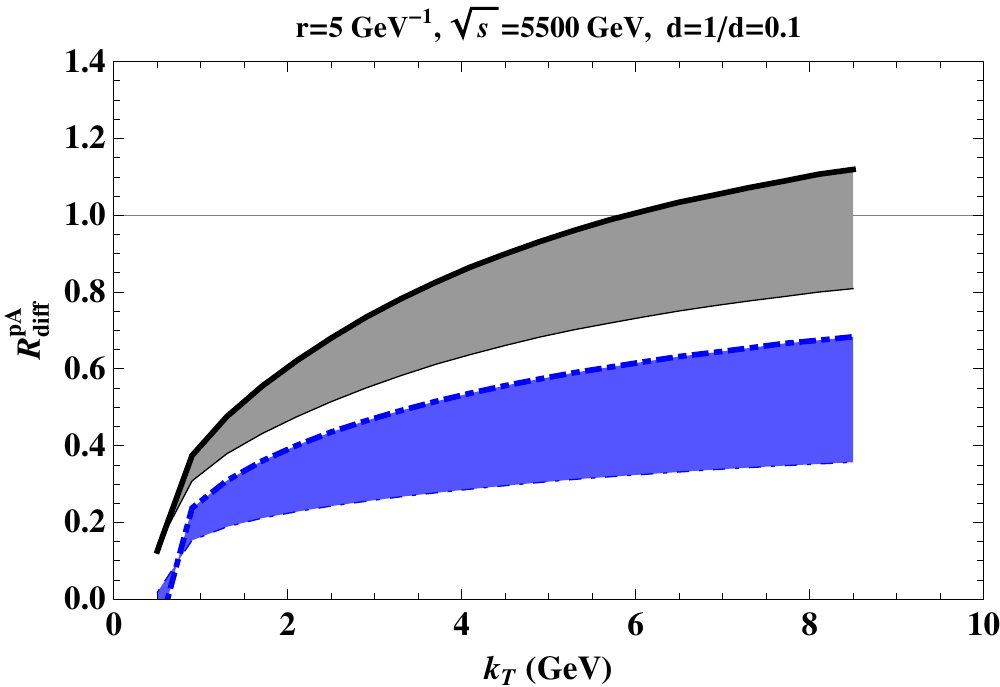}\\
   (a) & (b) 
\end{tabular}
\caption{Effect of diffusion in the dipole sizes on the diffractive pion production at LHC for  two characteristic sizes of proton (a) 0.2 fm and (b) 1 fm. Upper line of the same type corresponds to $d=1$, the lower one -- $d=0.1$.  Lines of different types correspond to different rapidities  (notations are the same as  in \fig{G2}). The effects of finite coherence length are neglected.}
\label{F2}
\end{figure}

The results of numerical calculations are exhibited in \fig{G1}--\fig{F2}. In \fig{G1} one can see that at RHIC $R^{pA}\sim 2-2.5$ at $y\simeq 0$   and $k_T>2$~GeV. This enhancement is a signature of a leading power correction, see \eq{j5}. As rapidity increases there are two important effects, which take place in the proton and nucleus wave functions: (i) spectrum of intermediate dipoles in a projectile proton shrinks as the rapidity interval available for the low-$x$ evolution in proton becomes  narrower, (ii) as $y$ increases, $x$ of gluon decreases causing stronger gluon saturation effect in the nucleus. Both effects lead to suppression of the nuclear modification factor. Gluon saturation in proton leads to the suppression law \eq{j10-1}. Of course, the effect of diffusion in  a proton is more pronounced for a proton with larger characteristic size, since in absence of the evolution effects (i.e.\ in the quasi-classical approximation) the cross section would be exponentially suppressed, see \eq{j24}. 

We further investigated the effect of diffusion by introducing the parameter $d$ in \eq{ndiff}. As has been repeatedly pointed out in this paper, it is the BFKL diffusion that makes the diffractive gluon production possible by generating intermediate dipoles of small size. Gluon saturation effects in proton may tame the BFKL diffusion \cite{Mueller:2002zm} leading to smaller effective diffusion coefficient. This effect is taken into account in \fig{F1} for RHIC and in \fig{F2}  for LHC. The shadow region in all figures demonstrates the difference in the nuclear modification factor between the cases of $d=1$ and $d=0.1$. Switching off the diffusion severely impacts the nuclear modification factor  at low energies/rapidities and for larger distances between the valence quarks in proton.

Unlike the nuclear modification factor for inclusive hadron production, which decreases as a function of rapidity and centrality at RHIC and reaches almost maximal possible suppression so that no significant additional suppression is expected at LHC, the diffractive hadron production shows a very interesting behavior even at LHC. This makes this process suitable for exploration of  different kinematic regions at the high energy frontier. We believe that it will be instrumental in unraveling the structure and dynamics of strong gluon fields.

%%%%%%%%%%%%%%%%%%%%%%%%%%%%%%%%
\acknowledgments
We would like to thank Dima Kharzeev, Yuri Kovchegov and J.-W.~Qiu for many informative discussions. 
The work of K.T. was supported in part by the U.S. Department of Energy under Grant No.\ DE-FG02-87ER40371. He would like to
thank RIKEN, BNL, and the U.S. Department of Energy (Contract No.\ DE-AC02-98CH10886) for providing facilities essential
for the completion of this work.

%%%%%%%%%%%%%%%%%%%%%%%%%%%%%%%%%%%%%

\end{document}